\documentclass[aps,prl,10pt,twocolumn,superscriptaddress]{revtex4}
\usepackage{mathrsfs}
\usepackage{epsfig, cancel, ulem}
\usepackage{latexsym}
\usepackage{natbib, comment}
\usepackage{url}
\usepackage{dcolumn}
\usepackage{color}
\usepackage{amsfonts,amssymb,amsmath, txfonts}
\usepackage{graphicx,epsfig}
\usepackage{psfrag}
\usepackage{subfigure}
\usepackage{hyperref}
\hypersetup{colorlinks=true}
\usepackage{mathtools}
\usepackage{enumitem}
\usepackage{float}
\usepackage[dvipsnames]{xcolor}
\usepackage{xcolor}
\hypersetup{ linktoc=all,
    colorlinks, linkcolor={brightpink},
    citecolor={blue}, urlcolor={blue}
}
\definecolor{rosy}{RGB}{230,235,252}
\definecolor{myframetitle}{RGB}{90,89,170}
\definecolor{myblocktitle}{RGB}{140,185,249}
\definecolor{mytitle}{RGB}{10,80,26}

\definecolor{darkgreen}{RGB}{27,130,45}
\definecolor{darkblue}{rgb}{0,0,0.3}
\definecolor{darkred}{rgb}{0.7,0,0}

\definecolor{light gray}{RGB}{220,220,220}
\definecolor{dark purple}{RGB}{108,0,217}
\definecolor{pink}{RGB}{190,20,100}
\definecolor{orang}{RGB}{193,63,0}
\definecolor{green}{RGB}{11,98,17}
\definecolor{darkpink}{RGB}{153,0,76}
\definecolor{bluegreen}{RGB}{0,102,102}
\definecolor{greenlagan}{RGB}{0,102,0}
\definecolor{redgreen}{RGB}{102,102,0}
\definecolor{Redgreen}{RGB}{153,76,0}
\definecolor{vividviolet}{rgb}{0.62, 0.0, 1.0}
\definecolor{amaranth}{rgb}{0.9, 0.17, 0.31}
\definecolor{palatinateblue}{rgb}{0.15, 0.23, 0.89}
\definecolor{brightpink}{rgb}{1.0, 0.0, 0.5}
\definecolor{cornflowerblue}{rgb}{0.39, 0.58, 0.93}
\definecolor{deepcarminepink}{rgb}{0.94, 0.19, 0.22}
\definecolor{radicalred}{rgb}{1.0, 0.21, 0.37}

\def\weff{w_{\tiny{\text{eff}}}}
\def\H0{{\text{H}\hspace*{-2.05mm}\text{H} 0\hspace*{-1.35mm}0\ }}

\def\be{\begin{equation}}
\def\ee{\end{equation}}
\def\beq{\begin{equation}}
\def\eeq{\end{equation}}
\def\bea{\begin{eqnarray}}
\def\eea{\end{eqnarray}}
\newcommand{\dd}{\textrm{d}}

\begin{document}

\title{Running Hubble Tension and a H0 Diagnostic }

\author{C. Krishnan} \email{chethan.krishnan@gmail.com}
\affiliation{Center for High Energy Physics, Indian Institute of Science, Bangalore 560012, India}
 \author{E. \'O Colg\'ain}\email{ocolgain.eoin@apctp.org}
 \affiliation{Asia Pacific Center for Theoretical Physics, Postech, Pohang 37673, Korea}
 \affiliation{Department of Physics, Postech, Pohang 37673, Korea}
\author{M. M. Sheikh-Jabbari}\email{shahin.s.jabbari@gmail.com}
\affiliation{School of Physics, Institute for Research in Fundamental Sciences (IPM), P.O.Box 19395-5531, Tehran, Iran}
\author{T. Yang}\email{tao.yang@apctp.org}
\affiliation{Asia Pacific Center for Theoretical Physics, Postech, Pohang 37673, Korea}

\begin{abstract}
Hubble tension is routinely presented as a mismatch between the Hubble constant $H_0$ determined locally and a value inferred from the flat $\Lambda$CDM cosmology. In essence, the tension boils down to a disagreement between two numbers. Here, assuming the tension is cosmological in origin, we predict that within flat $\Lambda$CDM there should be other inferred values of $H_0$, and that a ``running of $H_0$ with redshift" can be expected.
These additional determinations of $H_0$ may be traced to a  difference between the effective equation of state (EoS) of the Universe within the Friedmann-Lema\^itre-Robertson-Walker (FLRW) cosmology framework and the current standard model. We introduce a diagnostic that flags such a running of $H_0$. 
\end{abstract}

\maketitle

\section{Introduction}
Hubble tension suggests that we may be seeing cracks in the flat $\Lambda$CDM cosmological model \cite{Verde:2019ivm, DiValentino:2020zio}. At the crux of the tension, one finds a significant disagreement between a local determination of the Hubble constant $H_0$ due to the SH0ES collaboration \cite{Riess} and the value inferred by the Planck team through an analysis of the Cosmic Microwave Background (CMB) within the flat $\Lambda$CDM framework \cite{Aghanim:2018eyx}. A host of recent other determinations of $H_0$ exist, some supporting the idea that there is a discrepancy  \cite{Wong:2019kwg, Huang:2019yhh, Pesce:2020xfe, Dhawan:2020xmp, Schombert:2020pxm}, while others caution that the tension may be a mirage \cite{Freedman:2019jwv, Birrer:2020tax, Efstathiou:2020wxn, Kim:2020gai}. The jury is still out. 

We will argue in this paper that, if Hubble tension is substantiated further by upcoming observations, and if it is cosmological in origin, then the inferred $H_0$ within the FLRW cosmology framework can run as a function of data redshift. Put bluntly, either $H_0$ is unique and there is no tension, or further determinations of $H_0$ with different values should be anticipated as new redshift ranges are elucidated by observation. Interestingly, the H0LiCOW collaboration have presented tentative results \cite{Wong:2019kwg, Millon:2019slk}, further supported by \cite{sliding-H0, Dainotti:2021pqg}, which tease a potential descending trend of $H_0$ with redshift below $z \sim 0.7$ \footnote{{Note that the H0LiCOW's result is  a descending  $H_0$ with \textit{lens redshift}, whereas the cosmological constraints come from the time-delay distance, which is a combination of both the lens and source redshifts.  Therefore, their descending trend is not necessarily of a cosmological origin, and also one may not rule out the possibility of it being due to systematic errors. We thank Ken Wong for this comment.}}. The current significance of this descending feature is in the $\sim 2 \, \sigma$ range. 

{Here we emphasise that the argument is general, and that running can in principle occur at all redshifts, even if it turns out to be insignificant at $z\lesssim O(1)$.} Nevertheless, the Dark Energy Spectroscopic Instrument (DESI) \cite{Aghamousa:2016zmz} provides a unique opportunity to confirm or eliminate the feature in the low-redshift regime. Ultimately, if running in $H_0$ is substantiated beyond existing observations \cite{Wong:2019kwg, Millon:2019slk, sliding-H0, Dainotti:2021pqg}, this leads further credence to the cosmological origin of Hubble tension. {Conversely, if we do not find evidence for a running $H_0$, a non-cosmological explanation for the results in \cite{Verde:2019ivm, DiValentino:2020zio} becomes more likely.} 

Our main observation is that such a running in $H_0$ can be naturally understood and formulated within the FLRW cosmology. Recall that within the FLRW set-up, the Friedmann equations 
\begin{subequations}\label{Friedmann-Eq}
\begin{align}
\label{eq1} H^2 =& \frac{\rho}{3}, \\
\label{eq2} (1+z) \frac{H'}{H} =&  \frac{3(\rho+ p)}{2\rho}:=\frac32(1+\weff), 
\end{align}
\end{subequations}
must hold. Here, all expressions are a function of redshift $z$, {prime} denotes a derivative w.r.t to $z$ and we have set $M_{pl} = c = 1$ for simplicity. In addition, $\weff=\weff(z)$ denotes the effective equation of state (EoS) of the corresponding FLRW Universe \footnote{Here we need not assume flat FLRW cosmology. The contribution of the curvature term may be included as $\rho_k=-3 P_k=k(1+z)^2$ in the right-hand-side of Friedmann equations \eqref{Friedmann-Eq}.}. 

Equation (\ref{Friedmann-Eq}) can be integrated to get
\be
\label{eq3}
H_0 = H(z) \exp \left( - \frac{3}{2} \int^{z}_0 \frac{1+ \weff (z')}{1+z'} \dd z' \right).
\ee
It is worth emphasising again that this all follows from the FLRW paradigm, namely the assumption that the Universe is homogeneous and isotropic and is governed by Einstein's equations. Observe that $H_0$ does not appear in the Friedmann equations \eqref{Friedmann-Eq}: it is a result of integrating the equations and is hence  {\it identically} a constant.
Nonetheless, in practice \eqref{eq3} carries non-trivial  information regarding $H_0$. Recall that in a cosmological setting $H_0$ is observationally  determined by extrapolating the $H(z)$ read from data at higher $z$ to $z=0$, after a particular cosmological model is chosen, or equivalently, a choice of $\weff(z)$ is made. We work with a general $\weff(z)$ in our discussion instead of a specific cosmological model, to emphasize that $H_0$ is a truly  model independent quantity \footnote{If we have a cosmological model consisting of $N$ component fluid with energy densities $\rho_i$ and EoS $w_i$, then $\weff(z)=\sum_{i=1}^N \Omega_i w_i$, where $\Omega_i=\rho_i/\rho$.}. Finally, note that we can also replace $H(z)$ on the RHS of (\ref{eq3}) by the inverse of a derivative of the luminosity $D_{L}(z)$ or angular diameter distance $D_{A}(z)$. It can be applied beyond observational Hubble data (OHD).


More concretely, (\ref{eq3}) can be viewed in two complementary ways:\\
\textbf{(I)} {It represents a direct comparison between observations, for example OHD,  which determine $H (z_i)$  at given (low) redshifts $z_i$, and a model specified by $\weff(z)$.} Together, these quantities can be used to {\it define} a $H_0(z_i)$ (namely $H_0$ inferred from the observation at $z_i$) via the RHS of \eqref{eq3}.
In practice,  one can take cosmic chronometers (CC) \cite{Jimenez:2001gg} and baryon acoustic oscillations (BAO) \cite{Eisenstein:2005su} as {frequently used} OHD. Viewed in this light, the holy grail of cosmology is to identify the underlying $\weff(z)$ so that $H_0$ remains $z_i$ independent and a constant at all redshifts. When using \eqref{eq3} to determine $H_0$ in terms of observations and models, we are tacitly assuming that this is true. 
\\
\textbf{(II)} One can alternatively compare two different models, model A and model B, which are respectively specified by effective EoS  $\weff^{(A)}(z)$ and $\weff^{(B)}(z)$.
{If model A is the assumption}, but observations prefer model B, 
then running in $H_0$ is guaranteed. In particular, one can show that the ratio between the Hubble constants will generically have some $z$-dependence: 
\be\label{H0A/H0B}
\frac{H_0^{(A)}}{H_0^{(B)}} = \exp \left( \frac{3}{2} \int^{z}_0 \frac{\Delta \weff(z')}{1+z'} \dd z' \right)
\ee
where $\Delta \weff(z) := \weff^{(B)}(z) - \weff^{(A)}(z)$. Note, only if $\Delta \weff(z) = 0$  \textit{for all redshifts},  one can be sure that the RHS is a constant. Observe that {$H(z)$ which is an observational quantity  drops out in the ratio} and  the RHS of \eqref{H0A/H0B} is a  quantity integrated from today $z=0$ to a given redshift $z$, so even if $\Delta \weff(z) \approx 0$ in some intermediate redshift range, there can be non-vanishing tension. 

{One may assume that model A is a cosmological model within FLRW setup which yields  the SH0ES result \cite{Riess} for $H_0^{(A)}$ and model B can be  flat $\Lambda$CDM which yields the Planck value \cite{Aghanim:2018eyx} for $H_0^{(B)}$. Eq.\eqref{H0A/H0B} clearly shows that the LHS cannot remain a constant and it should run. Note that our argument applies to all approaches to resolving Hubble tension, for example \cite{ Bernal:2016gxb, DiValentino:2016hlg, Freedman:2017yms, Sola:2017znb, Yang:2018qmz, Li:2019yem, Agrawal:2019dlm, Vagnozzi:2019ezj, Keeley:2019esp, Pan:2019hac, Choi:2019jck, Akarsu:2019hmw, Sakstein:2019fmf, DAgostino:2020dhv, Alestas:2020mvb, DeFelice:2020sdq, Heisenberg:2020xak}, which change the EoS at some ranges of redshift.}  Note also that since the RHS of (\ref{H0A/H0B}) is integrated, one cannot infer $\Delta w_{\textrm{eff} } (z)$ uniquely from two discrepant $H_0$ values.

Thus, we see that a running $H_0$ is natural if we are comparing observations or models against (other) models. Constant $H_0$ represents the special case where the assumed model is correct. Let us further note from \eqref{H0A/H0B} that we have an explicit expression for the tension between $H_0$ at $z=0$ and that inferred from some finite $z$ -- it is the integral on the RHS, measuring the \textit{accumulated error} in the model equation of state.

We stress that the integrals in \eqref{eq3} or \eqref{H0A/H0B} can be explicitly performed once a cosmological model is chosen, e.g.  for flat $\Lambda$CDM model, one gets 
\be\label{LCDM-E(z)}
\exp\left(3\int^{z}_0 \frac{1+\weff(z')}{1+z'} \dd z'\right)=1- \Omega_{m0} + \Omega_{m 0} (1+z)^3,
\ee
where $\Omega_{m0}$ is the relative matter density at the present epoch. The integrated form of this equation shows that once a model is chosen, the accumulated error in the integral can be traded for the error in model parameters.

Since the Hubble constant of the present epoch is tautologically a constant in an FLRW universe, a ``running $H_0$" can be confusing. So let us summarise: in order to reconstruct $H_0$ from $H(z)$ data at some redshift, one must use the RHS of \eqref{eq3}. But to do that we need a model, as defined by $\weff(z)$, and this means \eqref{eq3} is no longer an identity if our model happens to be wrong. The key point here is that the only way one can compare the values of $H_0$ inferred from observations at two (sufficiently) different redshifts, is {via an assumption on $\weff(z)$, or in other words,} a model.

\section{A new diagnostic for \texorpdfstring{$\Lambda$CDM}{}} 

$H_0$ tension and a potential running $H_0$ can be sharply formulated recalling the viewpoint \textbf{(I)} discussed above by benchmarking the flat $\Lambda$CDM model against observational determinations of $H(z_i)$. \textit{A priori}, since the true EoS of the Universe is unknown, but is believed to correspond to flat $\Lambda$CDM to first approximation, there is no guarantee that the LHS of (\ref{eq3}) is a constant. We can hence use this fact to specify possible deviations from the flat $\Lambda$CDM model. This may be achieved through the $H_0$ diagnostic $\H0$:
\be
\label{H0}
\H0 (z) := \frac{H(z)}{\sqrt{1- \Omega_{m0} + \Omega_{m 0} (1+z)^3}}. 
\ee
As explained, $H(z)$ follows from OHD, for example CC or BAO, but the denominator requires a little explanation.  

Since we are primarily interested in low redshift observations, we have neglected radiation and neutrinos, but otherwise the denominator corresponds to the flat $\Lambda$CDM model. We propose that the denominator can be fixed by sampling $\Omega_{m0}$ directly from the Markov Chain Monte Carlo (MCMC) chains from the Planck mission \cite{Aghanim:2018eyx}. This in effect fixes the RHS so that it is only a function of redshift, yet allows one to account for errors in $\Omega_{m0}$. Alternatively, one can employ supernovae to determine the denominator and this should be attractive in coming years as WFIRST \cite{Spergel:2015sza} is expected to increase the number of supernovae by two orders of magnitude. Moreover, employing supernovae allows one to by-pass CMB and define an exclusively low-redshift diagnostic. Here we first adopt Planck values for $\Omega_{m0}$, since they provide the most constrained definition of the standard model. 

The $\H0 (z)$ \eqref{H0}, evaluated as discussed above, provides us with a null hypothesis test for the Planck $\Lambda$CDM model; if $\H0 (z)$ is not a constant within error bars, then Planck-$\Lambda$CDM model needs modification.

{\section{Illustrative Examples}} 

\begin{figure}[h]
\centering
\includegraphics[angle=0,width=80mm]{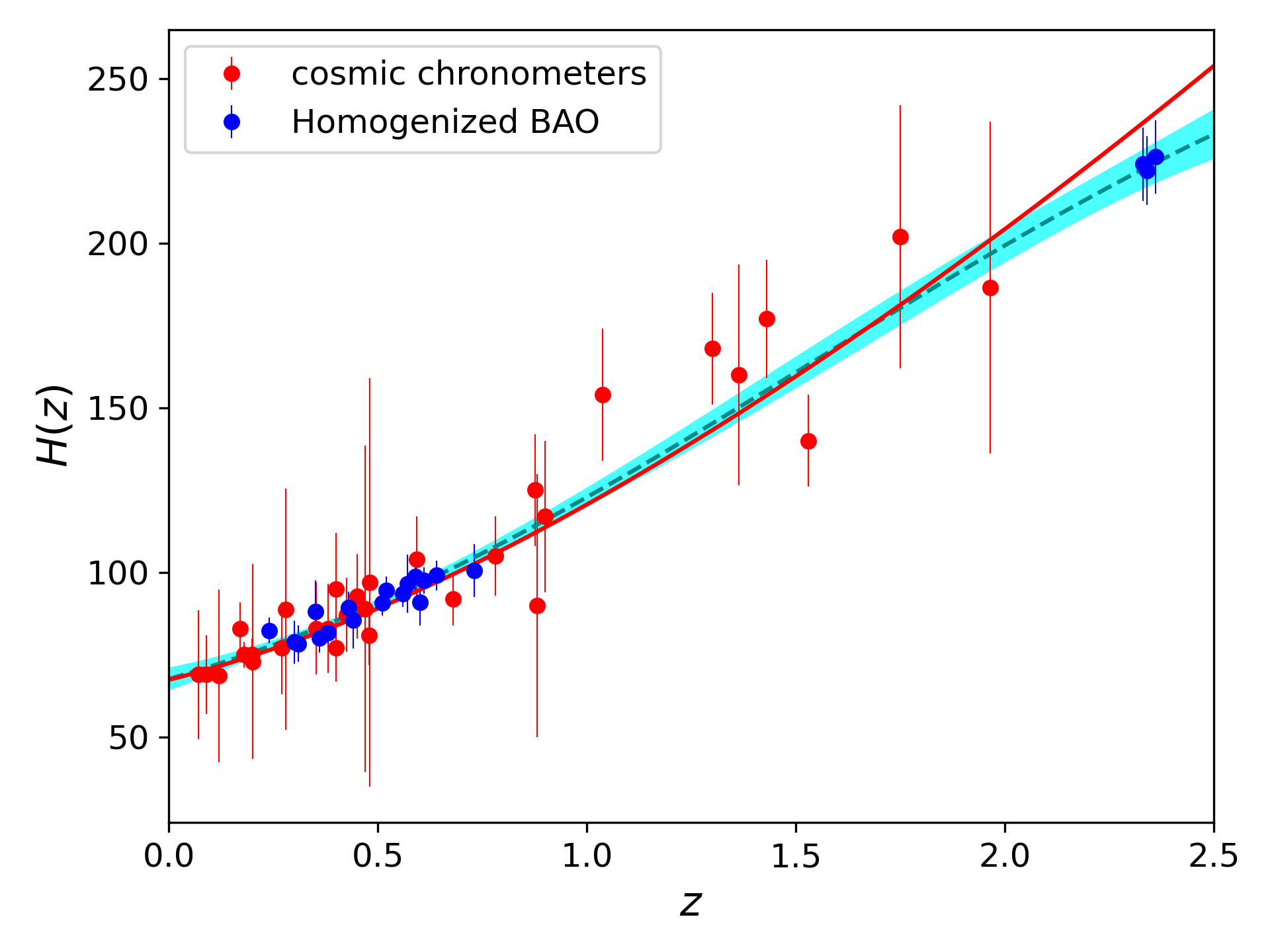} \\
\caption{The reconstructed Hubble parameter from the CC and BAO data. We illustrate the Planck-$\Lambda$CDM cosmology (red line) for comparison.  }
\label{fig1}
\end{figure}

{Let us work through some examples to see how the above discussions can be used in practice.} With an eye on imminent releases from DESI \cite{Aghamousa:2016zmz}, we will present results based on current data and future forecasted data. Concretely, we make use of the homogenised BAO data in Table 2 of ref. \cite{Magana:2017nfs} and following \cite{Shafieloo:2012ht, Seikel:2012uu}, we employ Gaussian Processes to reconstruct the Hubble parameter. Since the BAO data is relatively sparse, we augment the BAO with the CC data compiled in ref. \cite{Moresco:2016mzx}. The reconstructed Hubble parameter and data are presented in FIG. \ref{fig1}.  Observe that there is a noticeable difference in the quality of CC and BAO data, but since GP reconstructs data, we have added the CC data to provide guidance to GP where the BAO data is sparse. Throughout we have utilised the Mat\'ern covariance function with $\nu = \frac{9}{2}$. In a recent study \cite{Colgain:2021ngq} it has been shown over a large number of mocks of the same data that the  errors in $H_0$ decrease as $\nu \rightarrow \infty$, but the difference between $\nu = \frac{5}{2}$ and $\nu = \infty$ (Gaussian kernel) is not so pronounced. In other words, $\nu = \frac{9}{2}$ is pretty representative. Thus, FIG. \ref{fig1} is an illustrative snapshot of the current status of the data. At higher redshifts, we see a clear deviation from flat $\Lambda$CDM based on Planck values that is driven by Lyman-$\alpha$ BAO.  

\begin{figure}[h]
\centering
\includegraphics[angle=0,width=80mm]{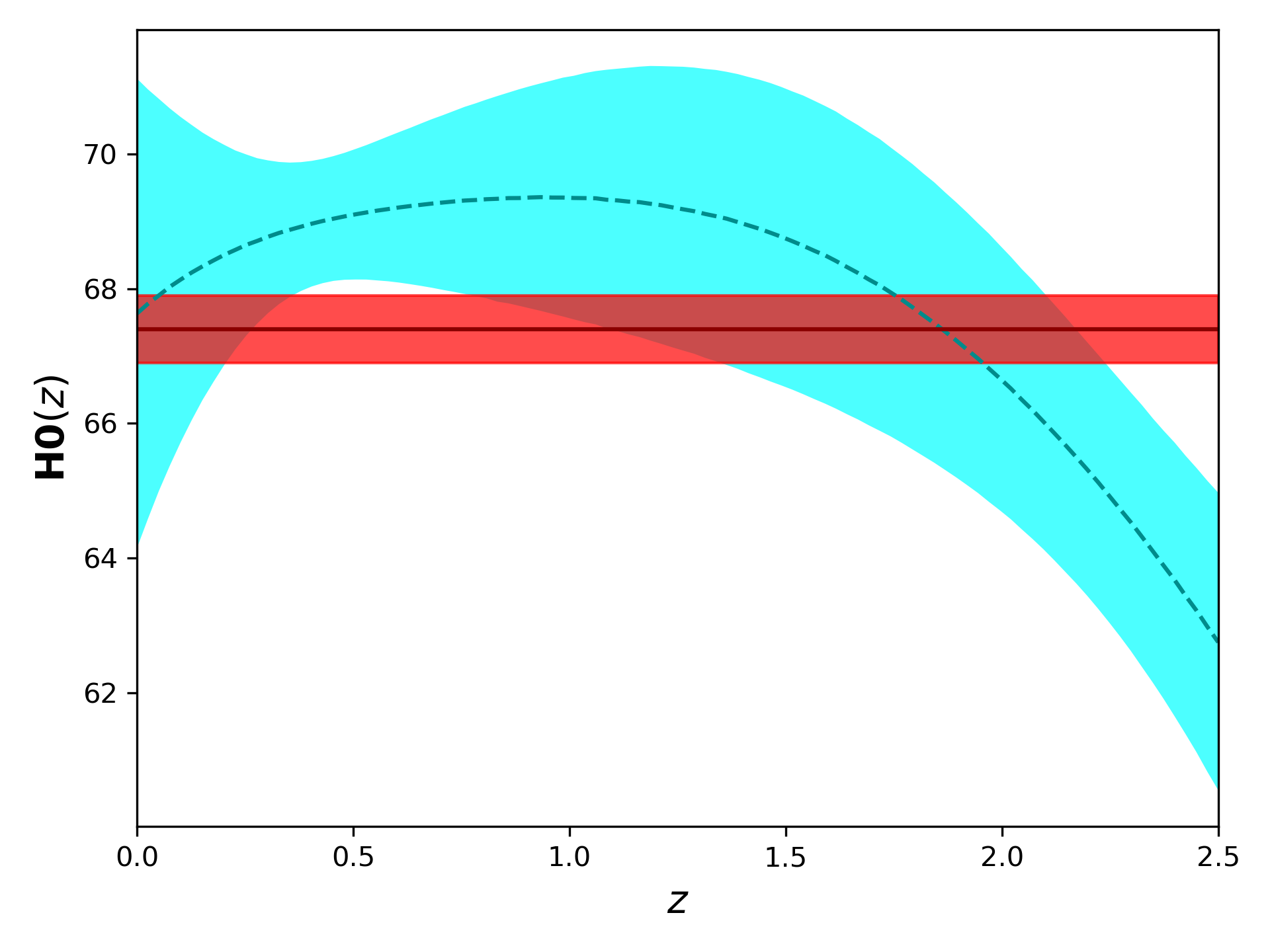} \\
\caption{The inferred value of our diagnostic ${\text{H}\hspace*{-1.8mm}\text{H} 0\hspace*{-1.2mm}0\ }$ in units of km/s/Mpc from the current CC and homogenised BAO data compared against the Planck-$\Lambda$CDM value in red.}
\label{fig2}
\end{figure}

This is in line with expectations: the Lyman-$\alpha$ BAO Hubble parameter is lower than Planck and this explains the dip in $\H0(z)$ at $z \approx 2.5$ in FIG. \ref{fig2}. In this case, the maximum deviation from the Planck value occurs at $z \approx 2.5$ and the statistical significance is $\sim 2 \, \sigma$ \footnote{{As discussed in \cite{Colgain:2021ngq}, the tension depends on the choice of kernel for the GP analysis and it can be enhanced by the reconstruction procedure.}}. If we restrict our attention to $z \lesssim 1$, where the data quality is better, one finds that the maximum deviation is $\sim 1.6 \, \sigma$ and this occurs at $z \approx 0.5$, as can be clearly seen from FIG. \ref{fig2}. In the range $0 < z < 1.5$, the data has a mild preference for a higher value of $\H0(z)$ than Planck. This could be attributed to the CC data, which favours a slightly higher value, e. g. \cite{Gomez-Valent:2018hwc}. Overall, given the status of current observations, there is little evidence for any deviation from flat $\Lambda$CDM. 

\begin{figure}[ht]
\centering
\includegraphics[angle=0,width=80mm]{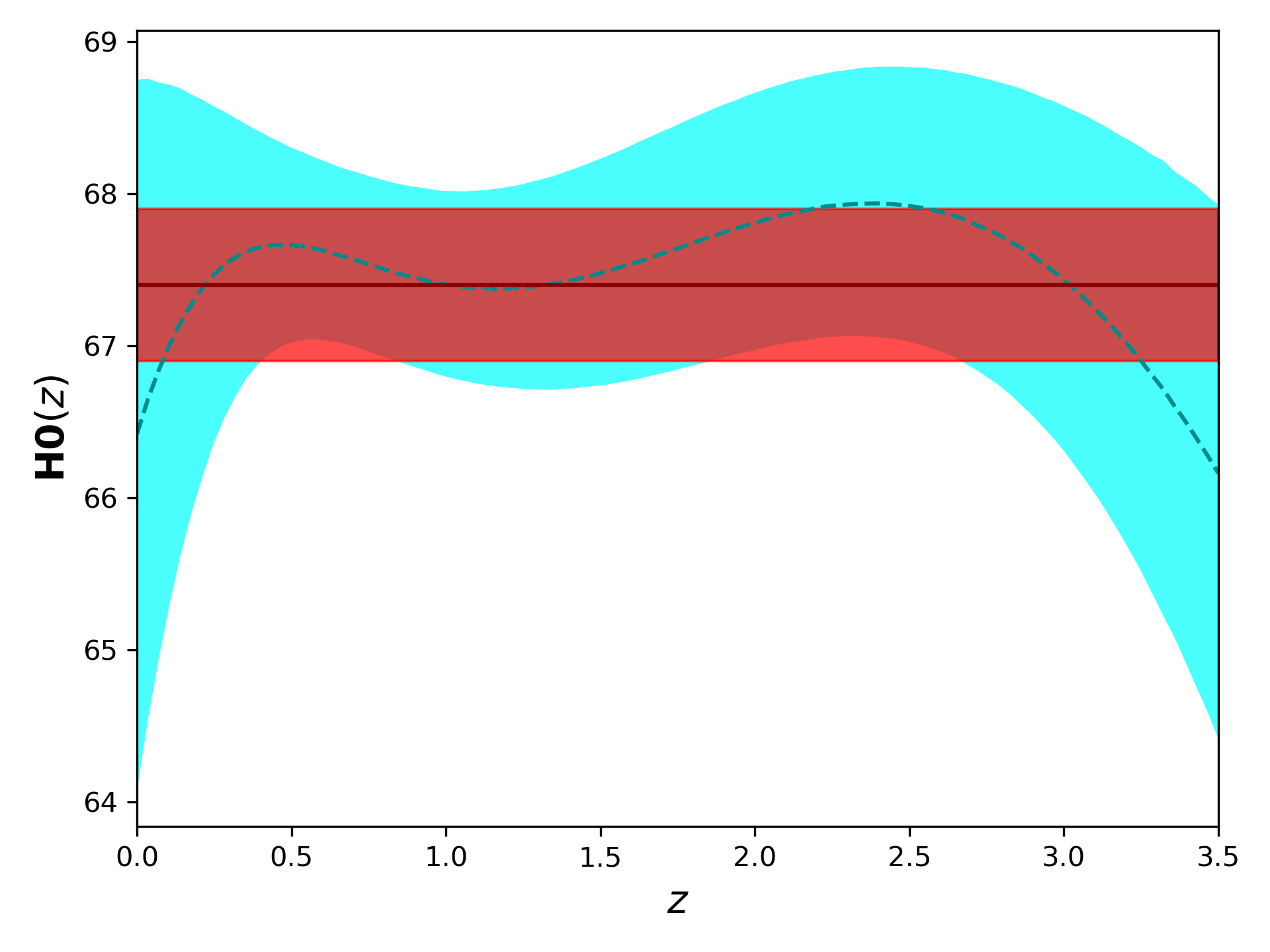} \\
\caption{The inferred value of our diagnostic ${\text{H}\hspace*{-1.8mm}\text{H} 0\hspace*{-1.2mm}0\ }$ in units of km/s/Mpc from the forecasted mock DESI BAO data compared against the Planck-$\Lambda$CDM value in red.}
\label{fig4}
\end{figure}

However, going forward one may be able to confirm or refute the idea that $H_0$ is running at low-redshift in flat $\Lambda$CDM through future DESI data \cite{Aghamousa:2016zmz}. In particular, we assume that the five-year survey covers 14,000 deg$^2$ (see Tables 2.3, 2.5 and 2.7 of \cite{Aghamousa:2016zmz}) and mock up data based on the flat $\Lambda$CDM model with canonical Planck values. In practice, one is mocking up 30 odd BAO data points in the redshift range $0.05 \leq z \leq 3.55$. In FIG. \ref{fig4} we have shown our $\H0$ diagnostic \eqref{H0} averaged over 100 mock realisations of the data. While there is no deviation from flat $\Lambda$CDM, and this is expected as flat $\Lambda$CDM was the basis for the mock, the key take-away message is that the errors in the $H(z)$ construction have contracted from FIG. \ref{fig2} to FIG. \ref{fig4}. In turn, this reduces the errors in our diagnostic. It is plausible that some running may be seen in future DESI releases. {That being said, it is worth emphasising that our diagnostic is equivalent to fitting a given model to binned data and extracting the $H_0$ determination in each bin. This alternative approach is essentially how the descending feature reported in \cite{sliding-H0} was identified. In future, it will be imperative to perform such consistency checks on cosmological parameters in order to elicit full confidence in a given model.}


\section*{Dynamical Dark Energy}
In recent years attempts have been made to reconstruct the dark energy sector directly from data through non-parametric techniques, whereby the EoS of dark energy is modeled through typically $30 \sim 40$ additional parameters relative to the base $\Lambda$CDM model \cite{Zhao:2017cud, Wang:2018fng}. In order to overcome the large uncertainties arising from additional parameters, one typically assumes that the parameters are correlated through a prior covariance matrix. The most recent analysis presents a $\sim 3.7 \, \sigma$ preference for dynamical dark energy \cite{Wang:2018fng}. 

Objectively, these non-parametric constructions have a large number of extra parameters, so they have additional freedom to fit datasets that may be discrepant. Some rigidity is provided by the assumption that the covariance matrix takes a particular form, but regardless, one typically encounters ``wiggles" as the variables oscillate around their flat $\Lambda$CDM values. Our aim here is to show that such wiggles will manifest themselves in differences in $H_0$ using our $\H0(z)$ diagnostic (\ref{H0}). Thus, if the dynamical dark energy claims of \cite{Wang:2018fng} are substantiated, one should expect to see pronounced wiggles in our diagnostic (\ref{H0}), or equivalently wiggles in $H_0$ within the flat $\Lambda$CDM model in different redshift bins. To date any deviations of low statistical significance are simply descending trends with no wiggles \cite{Wong:2019kwg, Millon:2019slk, sliding-H0, Dainotti:2021pqg}. In the big picture, not seeing these wiggles would ultimately cast doubts on the methodology of \cite{Zhao:2017cud, Wang:2018fng}.

\begin{figure}[h]
\centering
\includegraphics[angle=0,width=80mm]{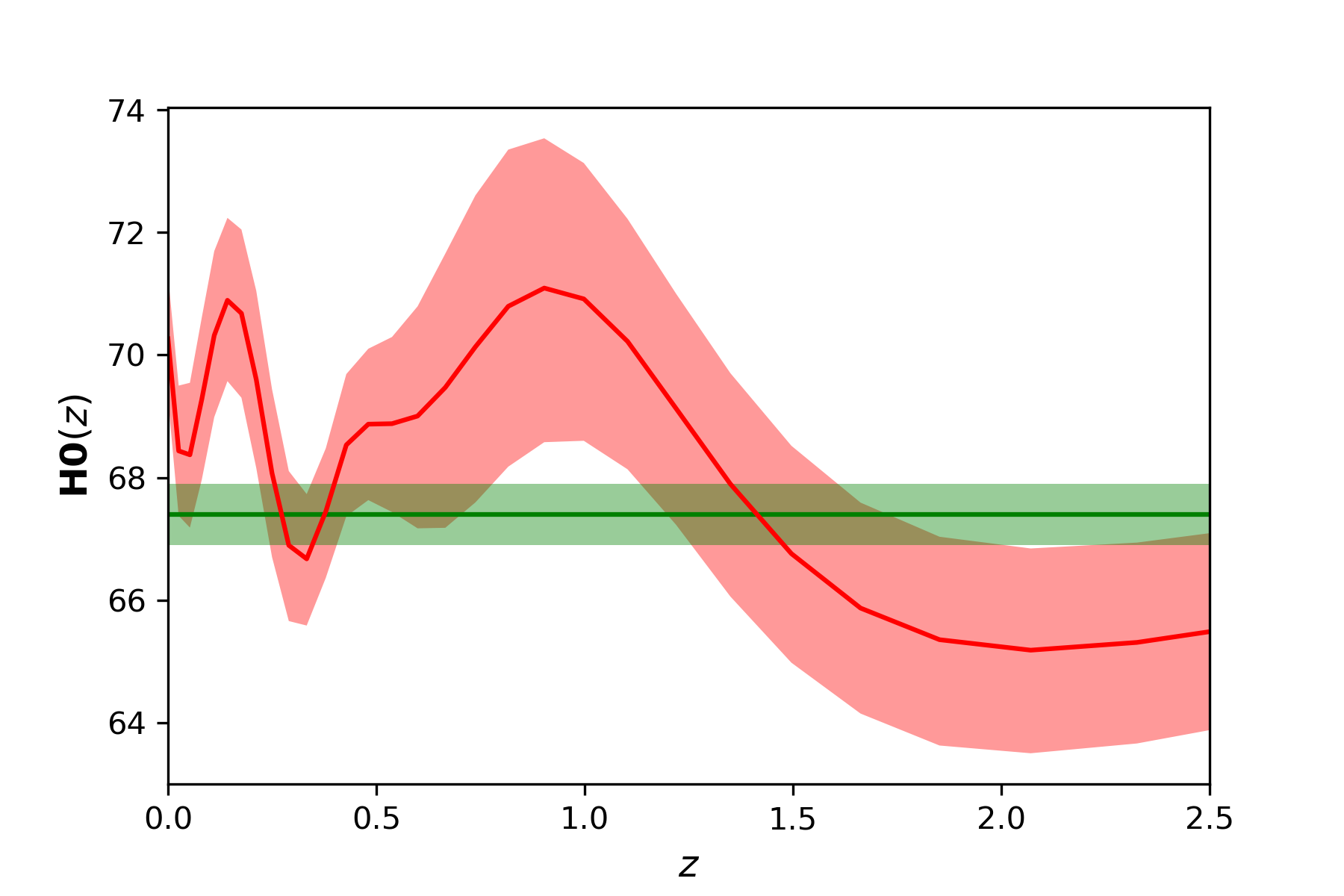} \\
\caption{The inferred value of ${\text{H}\hspace*{-1.8mm}\text{H} 0\hspace*{-1.2mm}0\ }$ from the $X$CDM reconstruction of \cite{Wang:2018fng} versus the Planck value.}
\label{fig5}
\end{figure}

Concretely, we start from the 41-dimensional $X$CDM model of ref. \cite{Wang:2018fng}, where in addition to $H_0$ and $\Omega_{m0}$, the authors allow for 39 uniform binned values of $X$ in the redshift range $z \in [0, 1000]$. In this context, $X$ is defined as the difference in the dark energy density $\rho_{\textrm{DE}}(z)$ at a given $z$ versus $z=0$, $X(z) \equiv \rho_{\textrm{DE}}(z)/\rho_{\textrm{DE}}(z=0)$. While this reconstruction may be questionable at higher redshifts, where data is sparse, here we focus on the lower redshift regime $z \lesssim 2.5$.  

\begin{figure}[h]
\centering
\includegraphics[angle=0,width=80mm]{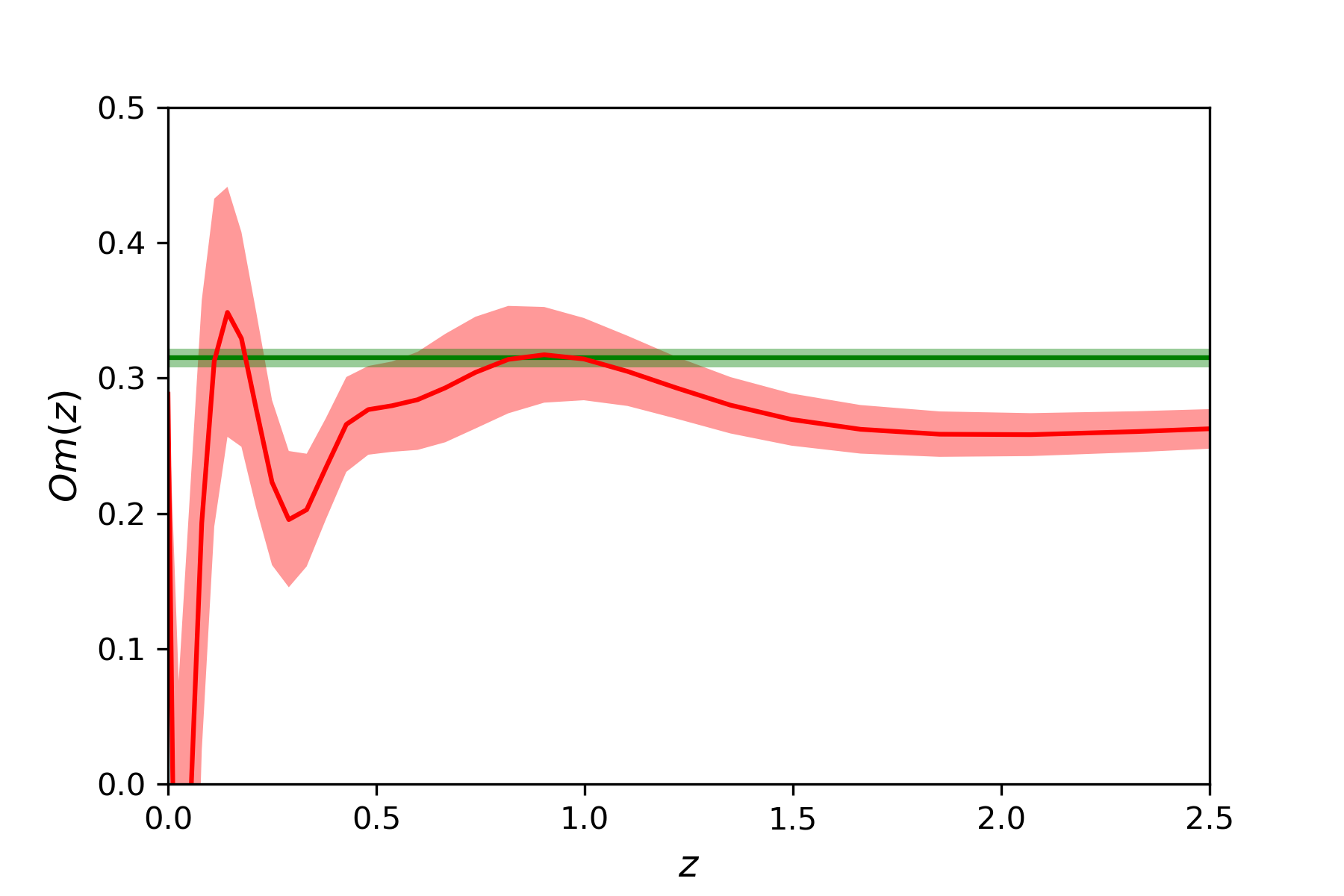} \\
\caption{The inferred value of $Om(z)$ from the $X$CDM reconstruction of  \cite{Wang:2018fng} versus the Planck value.}
\label{fig6}
\end{figure}

Indeed, the wiggles in $X$ around $X=1$ (flat $\Lambda$CDM) can be recast in terms of our diagnostic as shown in FIG. \ref{fig5}. Here, using the covariance matrix and mean values from \cite{Wang:2018fng}, we have generated a long chain of tuples $(X_i, H_0, \Omega_{m0})$, which can be converted into a chain of $H(z_i)$ determinations and restricted below $z \lesssim 2.5$. Dividing through as in \eqref{H0} with  $\Omega_{m0}$ from the Planck MCMC chains \cite{Aghanim:2018eyx}, we get the confidence intervals in FIG. \ref{fig5}. Taking into account the errors from the Planck determination of $H_0$, the maximum deviation occurs at redshift $ z \approx 0.16$ and the statistical significance is $2.5 \, \sigma$.  Interestingly, this is more or less the redshift where we see a deviation from flat $\Lambda$CDM in the Pantheon supernovae dataset \cite{Colgain:2019pck, mvp, Camarena:2019moy}, but here Wang et al. have employed the JLA dataset instead \cite{Betoule:2014frx}. The wiggles in FIG. \ref{fig5} are similar to Figure 2 of \cite{Wang:2018fng}, but there the data used is intrinsically low redshift, whereas our diagnostic comprises Planck MCMC chains: the diagnostic mixes low and high redshifts. One also observes wiggles in matter density $\Omega_{m0}$ in the Pantheon dataset when it is binned according to redshift \cite{Kazantzidis:2019dvk, Kazantzidis:2020xta, Kazantzidis:2020tko}. This observation is backed up by the compressed $E(z)$ data reported originally in \cite{Riess:2017lxs}. Moreover, at lower redshift $z \lesssim 0.7$, we see some indication of a descending $H_0$ with redshift in line with the findings of \cite{Wong:2019kwg, Millon:2019slk, sliding-H0}. The bump and decay can be attributed to Lyman-$\alpha$ BAO at $z= 2.34$ \cite{Delubac:2014aqe} and $z = 2.36$ \cite{Font-Ribera:2013wce}. 

For comparison, we also illustrate the $Om(z)$ diagnostic \cite{Sahni:2008xx, Zunckel:2008ti} in FIG. \ref{fig6}. Note that there is a pronounced dip just before $z=0$. While this dip is driven by the data, primarily a SH0ES prior on $H_0$, the construction of \cite{Wang:2018fng} demands that $X(z=0)=1$, so that $Om(z=0) = \Omega_{m0} = 0.288 \pm 0.008$. Deviations from the Planck result are once again evident and the significance exceeds $3 \, \sigma$ beyond $z \approx 1.85$. Curiously, this deviation is larger than the $\sim 1.7 \, \sigma$ deviation from Planck-$\Lambda$CDM that is usually attributed to Lyman-$\alpha$ BAO in the same redshift range. It would be interesting to revisit this study in future as more data becomes available. 

\section*{Comments On $w$CDM}
Our diagnostic can easily be extended to the $w$CDM model or further generalisations. Before doing this, let us recognise that the motivation for doing so may not be so great. First, Hubble tension is a $\sim 4 \sigma$ tension in the context of flat $\Lambda$CDM. By adding additional free parameters one can reduce the tension by inflating the errors. Moreover, if one believes the analysis in \cite{DAmico:2020kxu}, $w$CDM is tightly constrained to $ w= -1.046^{+0.055}_{-0.052}$ even without CMB data. Nevertheless, it is a valid exercise and may be instructive. 

To get the new diagnostic, one just needs to evaluate the RHS of (\ref{eq3}) for $w$CDM. Doing so, one finds
\be
\H0 (z) := \frac{H(z)}{\sqrt{(1- \Omega_{m0}) (1+z)^{3(1+w)} + \Omega_{m 0} (1+z)^3}}. 
\ee
Once again, one can extract $\Omega_{m0}$ and $w$ from the Planck MCMC chains and this fixes the RHS to be a function of $z$. Adding curvature is an immediate generalisation.

\begin{figure}[h]
\centering
\includegraphics[angle=0,width=80mm]{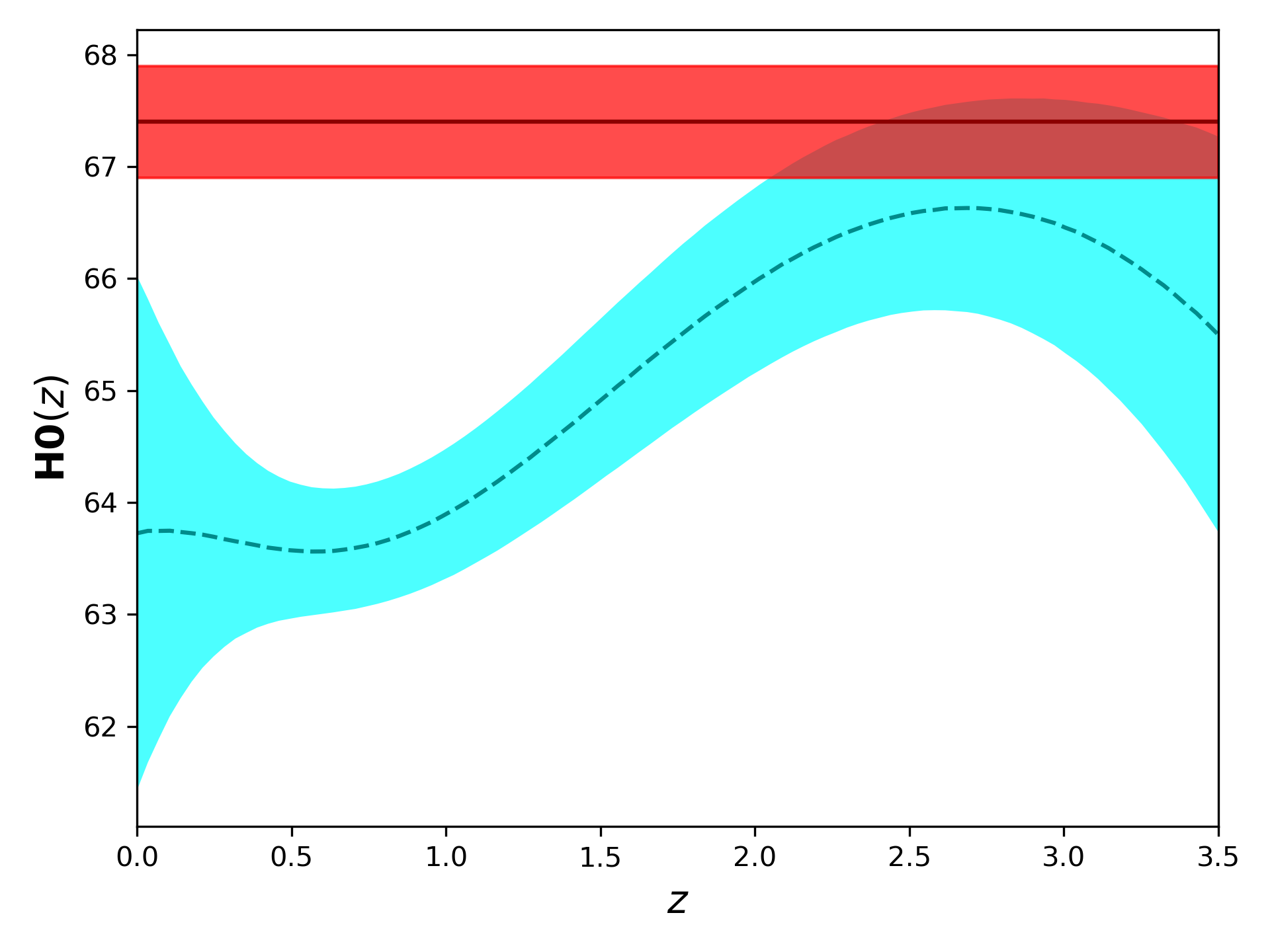} \\
\caption{Running of $H_0$ assuming flat $\Lambda$CDM (\ref{H0}) based on data mocked up as $w$CDM with $w = -1.3$. }
\label{fig7}
\end{figure}

Now that we have introduced the $w$CDM model, we can make one further comment. In \cite{Vagnozzi:2019ezj} $w \approx -1.3$ is proposed as a resolution to Hubble tension. Of course, this is at odds with \cite{DAmico:2020kxu}, but let us leave this aside for the moment. We can ask how would $H_0$ evolve if $w \approx -1.3$ and one assumed flat $\Lambda$CDM? One can answer this by mocking up $w$CDM data in the DESI forecasted range $0.05 \leq z \leq 3.55$ with the Planck values for $H_0$ and $\Omega_{m0}$, while setting $w = -1.3$. Here the assumed value of $H_0$ is not so important, simply the trend in $H_0$ captured by our diagnostic is of interest. This is somewhat close to the viewpoint formulated in item \textbf{(II)} and in \eqref{H0A/H0B}. In this case, one finds 
\begin{eqnarray}
\H0(z) &=& H_0 \frac{\sqrt{(1- \Omega_{m0}) (1+z)^{3(1+w)} + \Omega_{m 0} (1+z)^3}}{\sqrt{1- \Omega_{m0} + \Omega_{m 0} (1+z)^3}}  \nonumber \\
 &=& H_0 \biggl( 1 + \frac{3}{2} (1- \Omega_{m0} ) (1+w) z \\
 &+& \frac{3}{8} (1- \Omega_{m0} ) (1+w) ( 1 + 3 \Omega_{m0} (w-3) + 3 w) z^2 \biggr) \nonumber \\
 &+& \dots \nonumber
\end{eqnarray}
where we have expanded the first two terms to highlight the initial trend. By plotting the above analytic expression for $\Omega_{m0} = 0.3$ and $w = -1.3$, one sees that the ratio initially decreases with $z$ until $z \approx 0.3$ before increasing beyond that redshift. This feature is clearly visible from the mean values (dashed line) in FIG. \ref{fig7} where once again we have averaged over 100 mock realisations. It is interesting to note that the wiggles in $\H0(z)$ undulate over much larger redshift ranges than the $X$CDM model of \cite{Wang:2018fng} (see FIG \ref{fig5}) and that there is a low value of $\H0(z)$ relative to the Planck value at $z=0$. This appears to be an artifact of extrapolating beyond the range of the data. Since GP is a data reconstruction/interpolation technique, it is questionable beyond the outermost data points. 

\section{Discussion}
The point of this short note is to stress that a running $H_0$ should not instill us (or the H0LiCOW collaboration!) with fear, since it is very natural within the FLRW paradigm provided Hubble tension is indeed cosmological in origin. In essence, if there are two discrepant values of $H_0$, one should be able to identify further values: only when one correctly identifies the model through its effective EoS can one be confident that $H_0$ is a constant. At the moment, flat $\Lambda$CDM represents our best guess, but this is being challenged by Hubble tension. Nevertheless, given that the true effective EoS of the Universe is unknown, this motivates our diagnostic $\H0 (z)$ as a means to uncover potential running behaviour, tentative signals of which may have already been reported \cite{Wong:2019kwg, Millon:2019slk, sliding-H0, Dainotti:2021pqg}. 

It should be stressed that we have not specified a redshift where the EoS differs from flat $\Lambda$CDM, and it is conceivable that some variation in the early Universe is required, {as argued in \cite{Mortsell:2018mfj, Lemos:2018smw, Knox:2019rjx}}, for example {\cite{Poulin:2018cxd, Kreisch:2019yzn, Agrawal:2019lmo, Niedermann:2019olb, Niedermann:2020dwg} (see \cite{Hill:2020osr,Ivanov:2020ril,DAmico:2020ods, Niedermann:2020qbw, Murgia:2020ryi, Smith:2020rxx, Jedamzik:2020zmd} for related discussion}). Given probes of the pre-CMB Universe are limited, this may preclude us from observing a running in $H_0$. {However, a competitive determination of the age of the Universe from globular clusters may yet force us into a late Universe modification \cite{Bernal:2021yli}.} Therefore, in light of upcoming late Universe experiments: DESI \cite{Aghamousa:2016zmz}; WFIRST \cite{Spergel:2015sza}; Euclid \cite{Laureijs:2011gra}; it is timely to eliminate a variation in the EoS from flat $\Lambda$CDM at low redshift and the ensuing running in $H_0$. This is where our diagnostic may come into its own and it is an important consistency check on the flat $\Lambda$CDM model that $H_0$ does not vary with redshift. {Conversely, as argued in \cite{sliding-H0}, a confirmed running $H_0$ would disfavour early Universe resolutions to Hubble tension.} 

In this paper we have focused on OHD, essentially CC and BAO. As is clear from FIG. \ref{fig2}, there are hints of some deviations from the Planck value $H_0 = 67.4 \pm 0.5$ km/s/Mpc \cite{Aghanim:2018eyx} in current data, which can largely be attributed to Lyman-$\alpha$ BAO observations. Using the Gaussian Process technique \cite{Shafieloo:2012ht, Seikel:2012uu} for a non-parametric reconstruction of $H(z)$, any deviation is at most in the $\sim 2 \, \sigma$ window. That being said, DESI data is imminent and in FIG. \ref{fig4} we have illustrated how the confidence intervals will contract using the same methodology and forecasted mock BAO data based on flat $\Lambda$CDM with canonical Planck values. 

Our diagnostic has some overlap with the $Om(z)$ diagnostic \cite{Sahni:2008xx, Zunckel:2008ti} (see also \cite{Shafieloo:2012rs, Sahni:2014ooa}), but there are a few differences worth highlighting. First, $Om(z)$ is a diagnostic specific to flat $\Lambda$CDM leading to a yes-no statement on whether data is consistent with flat $\Lambda$CDM or not. In contrast (\ref{eq3}) {applies to any model}, a feature captured in the effective EoS. Secondly,
as we will argue in the appendix, $Om(z)$ is a less sensitive diagnostic than $\H0(z)$ at low redshifts. Thirdly, the strength of $Om(z)$ is that it does not depend on model parameters \footnote{Note that it depends on $H_0$ which is a direct astrophysical observable. In this respect $H_0$ is different from ``true'' model parameters, e.g. like $\Omega_{m0}$.}, whereas since our goal is to explicitly falsify  specific models, \eqref{eq3} and \eqref{H0} do depend on model parameters. In an era of late-cosmology dominated by Hubble tension, a simple diagnostic along the lines of (\ref{H0}) may be long overdue. 

Finally, one last comment is warranted. While the current CC and BAO data at best provides a hint of some running in $H_0$, nothing more nothing less, there are studies of dynamical dark energy where ``wiggles" in the EoS  are favoured over the constant dark energy EoS $w = -1$ at $3.7 \, \sigma$ \cite{Zhao:2017cud,Wang:2018fng}. Returning to \eqref{H0A/H0B}, one can compare the Planck value for $H_0$ directly to the value inferred from the difference in the EoS. This leads to a running in $H_0$, which can be captured by our diagnostic \eqref{H0} (see appendix). The key take-home here is that wiggles in the EoS will manifest themselves in differences in $H_0$, {when contrasted with flat $\Lambda$CDM}.

\section*{Acknowledgements}
 We thank Yuting Wang and Gong-Bo Zhao for sharing their mean values and covariance matrix from ref. \cite{Wang:2018fng}.  We are grateful to Anjan Sen and Ruchika for discussions on {the} running $H_0$ idea and collaboration on {the} earlier paper \cite{sliding-H0}. We thank Stephen Appleby, Lavrentios Kazantzidis, Benjamin L'Huillier, Varun Sahni, Arman Shafieloo, Alexei Starobinsky, Yuting Wang, Kenneth Wong and Gong-Bo Zhao for discussion and/or comments on  the draft. Note, acknowledgments do not imply concordance in opinion. This work was supported in part by the Korea Ministry of Science, ICT \& Future Planning, Gyeongsangbuk-do and Pohang City. MMShJ acknowledges the support by INSF grant No 950124 and Saramadan grant No. ISEF/M/98204.

\appendix 

\section*{Other diagnostics for \texorpdfstring{$\Lambda$CDM}{}}

Here we discussed the $\H0$ diagnostic \eqref{H0} and  there has been another diagnostic discussed in the literature, the $Om(z)$ diagnostic \cite{Sahni:2008xx, Zunckel:2008ti},
\be\label{Om}
Om(z) := \frac{E^2(z) -1}{(1+z)^3-1}, 
\ee
where $E(z) := H(z)/H_0$ is the normalised Hubble parameter. When $H(z)$ is exactly the one predicted by the flat $\Lambda$CDM, $Om(z)=\Omega_{m0}=const.$ and any $z$ dependence in $Om(z)$ signals a deviation from flat $\Lambda$CDM.

It is inevitable that  $Om(z)$  has large error bars once evaluated using low redshift data \cite{Sahni:2014ooa}. To see this, note that we can Taylor expand any cosmology $E(z) = 1 + (1+q_0)z + O(z^2)$, where $q_0$ is the deceleration parameter. Thus, at low redshift, $Om(z)$ reduces to $Om(z) = \frac{2}{3} (1+q_0) + O(z)$. Now, since $q_0$ is subleading in $z$ relative to $H_0$, one has to go to suitably high $z$ to determine it to some degree of precision. As a result, if there is a deviation at sufficiently low redshift, we can expect $Om(z)$ to be insensitive. This partially motivates our new diagnostic. 

One may also wonder if one can construct other similar  diagnostics within $\Lambda$CDM. There are three parameters $H_0, \Omega_\Lambda,\Omega_{m0}$ in $\Lambda$CDM which are related by two equations,
\be
H(z)^2=H_0^2 \left(\Omega_{m0}(1+z)^3+ \Omega_\Lambda\right),\qquad \Omega_\Lambda+\Omega_{m0}=1. 
\ee
One can hence think of three such diagnostics:\\
(1) $Om(z)$ \eqref{Om} which is based on eliminating $H_0, \Omega_\Lambda$ from the above equations and  the fact that for \texorpdfstring{$\Lambda$CDM}{}
$E(z)^2-1=\Omega_{m0}\left((1+z)^3-1\right)$.\\
(2) Our $\H0$ \eqref{H0} which stems from constancy of $H_0$ and\\
(3) $\mathbb{DE}(z)$, which stems from eliminating $H_0, \Omega_{m0}$ and writing $(1+z)^3-E(z)^2=\Omega_{\Lambda}\left((1+z)^3-1\right)$.
One can, however, readily see that $\mathbb{DE}(z)=1-Om(z)$ and hence it is not independent of $Om(z)$.

There are therefore only two such diagnostics. $Om(z)$ is fitter to capture deviations in the energy budget of the universe from that of flat $\Lambda$CDM, whereas $\H0$ is more apt to flag an ``integrated'' difference of a model (conveniently chosen as the standard flat $\Lambda$CDM cosmology) with the data or another model, within a range of redshifts.

What $Om(z)$ really checks for is if $\Omega_{m0}$ is \textit{ some} constant. The diagnostic \eqref{H0} on the other hand works once the model is completely specified by providing  specific values for the cosmological parameters before it can return us a result for the null test. The reason behind this is that $\Omega_{m0}$ is a parameter in the model, while $H_0$ is an integration constant. In this sense, the $H_0$ diagnostic is a more resolved diagnostic of models. Of course, since data precision is a bottleneck, this point is somewhat academic at the moment.

\end{document}